
\input harvmac

\Title{\vbox{\hbox{HUTP-95/A022, IC/95/115, MRI-PHY/12/95 }
\hbox{\tt hep-th/9506122}}}
{$c=1$ String as the Topological Theory of the Conifold}
\bigskip
\centerline{Debashis Ghoshal\foot{E-mail: ghoshal@theory.tifr.res.in.}}
\bigskip\centerline{\it International Centre for Theoretical
Physics}
\centerline{\it Trieste 34100, Italy}
\smallskip\centerline{\&}
\centerline{\it Mehta Research Institute of Mathematics \&\
Mathematical Physics\foot{Permanent address.}}
\centerline{\it 10 Kasturba Gandhi Road, Allahabad 211 002, India}

\vskip .1in
\centerline{and}

\vskip .1in

\centerline{Cumrun Vafa\foot{E-mail: vafa@string.harvard.edu.}}
\bigskip\centerline{\it Lyman Laboratory of Physics}
\centerline{\it Harvard University}\centerline{\it Cambridge,
MA 02138, USA}

\vskip .3in
We show that the non-critical $c=1$ string at the self-dual
radius is equivalent to topological strings based
on the deformation of the conifold singularity of Calabi-Yau
threefolds. The Penner sum giving the genus expansion of the
free energy of the $c=1$ string theory at the self-dual
radius therefore gives the
universal behaviour of the topological
partition function of a Calabi-Yau threefold near a conifold point.

\Date{\it {June 1995}}

String theories compactified on Calabi-Yau manifolds are well
motivated. These theories have also been extensively studied
and we have been rewarded by some rather novel and unexpected
features, which, it is believed, will enrich our understanding
of the non-perturbative features of string theory. Particularly
intriguing is the occurence of a generic class of singularities,
called conifolds \ref\conifold{ P. Candelas, A.M. Dale, C.A. L\"utken, and R.
Schimmrigk,
``Complete Intersection Calabi--Yau Manifolds'', {\it Nucl. Phys.} {\bf
B298} (1988)
493--525;  P.S. Green and T.  H\"ubsch,
``Possible Phase Transitions Among Calabi--Yau Compactifications'',
{\it Phys. Rev. Lett.} 61 (1988) 1163--1166;
``Connecting Moduli Spaces of Calabi--Yau Threefolds'', {\it Comm. Math.
Phys.} 119 (1988) 431--441;
P. Candelas, P.S. Green, and T.
H\"ubsch,
``Finite Distance Between Distinct Calabi--Yau Manifolds'',
{\it Phys. Rev. Lett.} 62 (1989) 1956--1959;
``Rolling Among Calabi--Yau Vacua'', Nucl. Phys.
 B330 (1990) 49--102;
 P. Candelas and X.C. de la Ossa,
``Comments on Conifolds'', {\it Nucl. Phys.} {\bf B342} (1990) 246--268.},
in the moduli spaces of Calabi-Yau threefolds.
At these points, the threefold develops a singularity where a
3-cycle shrinks to zero size. This would lead to some apparently
singular physical behaviour. A beautiful resolution to this
singularity problem (at least for the type II strings) has
recently been proposed in
\ref\strom{A.\ Strominger, ``Massless black holes and conifolds
in string theory'', hep-th/9504090.},
where it is shown that the problem
has its origin in our understanding of the low energy effective
field theory near the conifold. Black holes that were previously
massive, become light and eventually massless as we hit the
singularity. This was used to
propose \ref\gms{B.\ Greene, D.\ Morrison and A.\ Strominger,
``Black hole condensation and the unification of string vacua'',
hep-th/9504145.}\
a smooth non-perturbative stringy process interpolating two
completely different Calabi-Yau threefolds
thereby changing the topology of the internal space.  A test
of this proposal was made in \ref\vfate{C.\ Vafa, ``A stringy test of
the fate of the conifold'', hep-th/9505023.}\ by studying
the topological partition function
of the Calabi-Yau near the conifold at one-loop.  It was observed
there that there should be some
universal features of the topological partition function
at one loop as a consequence of the
proposal in \strom . This was verified in some examples for which the
topological partition function at one loop was computed.

In this paper we will demonstrate a further generalization of this universality
by arguing that the $c=1$ string theory
at the self-dual radius is equivalent to the topological theory
(of the B-model) corresponding to the deformation near a
conifold. A consequence  of this fact is that the genus expansion
of the free energy of the solvable $c=1$ string theory (at the
self-dual radius) captures the universal singular behaviour of
the free energy of topological partition function
of threefolds near a conifold point. We will illustrate this
by the example of the degenerating quintic.
Fortunately, our task is greatly
simplified by the fact that most of the relevant observations
have been made in the literature. In the following, we will
recall the necessary  facts and show how they lead to the
proposed link.

Recall that a conifold is a singular three (complex) dimensional
Calabi-Yau manifold. By varying the moduli that deform the complex
structure of the threefold, one can, by a finite change of the complex
structure, reach a Calabi-Yau space which has a singularity
\conifold . Near the singularity, the local structure of the
degenerating threefold can be described by the quadric \conifold :
$$x^2+y^2+z^2+t^2=\mu$$
where $x,y,z,t$ are coordinates of ${\bf C}^4$.  As the parameter
$\mu \rightarrow 0$ the quadric
develops a conical singularity at the origin. It is convenient for
our purpose to make a linear redefinition of the local coordinates
so that the quadric takes the following form
\eqn\coni{x_1x_2-x_3x_4=\mu}
This change of variable is standard, and lets us infer that
the base of the cone is $S^3\times S^2$. Turning on a non-zero
$\mu$ amounts to smoothening the singularity by replacing the
apex of the cone by an $S^3$ of `size' $\mu$. (There is
another way to desingularize
the conifold, called the `small resolution'. In this process
the origin is replaced by a ${\bf CP}^1\sim S^2$. We will however
focus our attention on the deformation by $\mu$.)

Let us now recall some aspects of the topological
theory of Calabi-Yau threefolds coupled to topological gravity.
In \ref\wit{E.\ Witten, ``On the structure of
the topological phase of two-dimensional gravity'', {\it Nucl.\ Phys.}
{\bf B340} (1990) 281.}\
it was shown that one can couple topological sigma models
to topological gravity, and in many ways, such theories
behave like ordinary non-critical matter coupled to 2d gravity.
Moreover it was noted that the topological sigma models based
on Calabi-Yau threefolds have a special property. For these
theories, the partition function is non-vanishing in all loops
without the need of any operator insertion. In this sense these
are critical theories, and the critical value of the central
charge $\hat c$ of the topological algebra is $\hat c=3$.

Topological sigma models on Calabi-Yau threefolds were studied in great
detail in \ref\bcov{M.\ Bershadsky, S.\ Cecotti, H.\ Ooguri and
C.\ Vafa, ``Kodaira-Spencer theory of gravity and exact results
for quantum string amplitudes'', {\it Comm.\ Math.\ Phys.}{\bf 165} (1994)
311.}\ with the discovery of the
importance of holomorphic/topological anomalies in such cases. Moreover
a field theory for the so called B-model of this topological theory was
constructed. Since it describes the behaviour of the moduli that are
sensitive to the deformation of the complex structure of the Calabi-Yau,
it is called the `Kodaira-Spencer theory of gravity'. The
K\" ahler moduli are frozen in the Kodaira-Spencer theory ---holomorphic
3-form preserving diffeomorphisms of the threefold are the symmetries
of the theory.

The topological algebra underlying any string background, in
particular minimal matter coupled to gravity were found
in \ref\gatosemi{B.\ Gato-Rivera and A.M. Semikhatov,``
Minimal models from W constrained hierarchies via the Kontsevich-Miwa
transform'', {\it Phys. Lett. }{\bf B288} (1992) 38.}\ref\bes{M. Bershadsky,
W. Lerche, D. Nemeschansky and N.P. Warner, ``Extended N=2 superconformal
structure of gravity and W-gravity coupled to matter'', {\it
Nucl. Phys.} {\bf B401} (1993) 304.}.
The interesting case of $c=1$ theory, that is one
free scalar compactified at the self-dual radius, coupled to gravity
was explored in \ref\muv{S.\ Mukhi and C.\ Vafa, ``Two-dimensional
black-hole as a topological coset model of c=1 string theory'',
{\it Nucl. Phys.} {\bf B407} (1993) 667.},
where the topological
$SL(2,R)_3/U(1)$ coset model \ref\witN{E.\ Witten, ``The N matrix
model and gauged WZW models'', {\it Nucl.\ Phys.}{\bf B371} (1992) 191.}\
was found to be equivalent to this background in the KPZ formalism.
This work also provides an explanation for the the observations of
\ref\dv{J.\ Distler and C.\ Vafa,``A critical matrix model at $c=1$'',
{\it Mod. Phys. Lett.} {\bf A6} (1991) 259.}\witN\
concerning the relation between
the $c=1$ string at self-dual radius to the Penner model. Moreover
it was noted that although the theory is traditionally interpreted
as a string moving in two dimensions,
the topological theory is critical --- it has the
same central charge as that of a Calabi-Yau threefold!

The simplest way to describe this topological
theory is in terms of a Landau-Ginzburg theory.
The topological Landau-Ginzburg model
in question is characterized by a singular superpotential
$W(X) = -\mu X^{-1}$\ref\cev{S.\ Cecotti and C.\ Vafa, as referred
to in \muv.}\ref\gmone{D.\ Ghoshal and S.\ Mukhi, ``Topological
Landau-Ginzburg model for two-dimensional string theory'', {\it
Nucl.\ Phys.}{\bf B425} (1994) 173.}\ref\hop{A.\ Hanany, Y.\ Oz and
R.\ Plesser,``Topological Landau-Ginzburg formulation and integrable
structure of 2-D string theory'',{\it Nucl. Phys.} {\bf B425} (1994) 150.}.
This model successfully reproduces the results
of matrix model at tree level \gmone\ and has an
integrable structure underlying it \hop.
Moreover it can also be used to compute the tachyon correlators
at higher loops \ref\gim{D.\ Ghoshal, C.\ Imbimbo and S.\ Mukhi,
``Toplogical 2D string theory: Higher genus amplitudes and
$W_\infty$ identities'', {\it Nucl.\ Phys.}{\bf B440} (1995) 355.}.

A step towards geometrization of this Landau-Ginzburg theory
was taken in \ref\gmtwo{D.\ Ghoshal and S.\ Mukhi,
``Landau-Ginzburg model for a critical topological string'', to appear
in the Proceedings of the International Colloquium on Modern
Quantum Field Theory, Eds.\ S.\ Das et al.}\ by noting
that one can relate it to a Calabi-Yau threefold by the usual
trick of adding extra quadratic terms to the superpotential.
Since the central charge is that of a threefold, (quadratic
terms do not contribute to the central charge), one needs four
extra fields, which together with the original variable $x$
make up for the coordinates of the ambient space. The superpotential
now takes the form
$$W(x) = -\mu x^{-1} + y_1^2 + y_2^2 + y_3^2 + y_4^2$$
For this equation to make sense the space in which it is embedded
should be a weighted projective space ${\bf WCP}^4_{-2,1,1,1,1}$.
Notice that since the degree of $W$ equals the sum of weights,
the manifold defined by $W=0$ satisfies the Calabi-Yau condition.
It turns out to be convenient for us to represent the modified
superpotential in an equivalent way by redefining variables
\eqn\lan{W=-\mu x^{-1}+x_1x_2 -x_3x_4}
Viewed in this geometric form, the locus $W=0$ is nothing but the
conifold! More precisely the Calabi-Yau phase of the $c=1$
string theory is the deformation of a threefold near a conifold
singularity. The quick way to see this is to go to the coordinate
patch
where $x\ne 0$, choose $x=1$ without any loss of generality, and
note that in this affine patch $W=0$ is equivalent to the
definition of the deformed conifold. The cosmological constant
$\mu$ in Eq.\coni\ plays the role of the complex modulus whose
vanishing signals the appearance of the conifold singularity.
Note that the coordinate patch with $x=0$ has an infinite potential
associated with it and thus it does not arise in the geometrical
description of this theory.
This correspondence can be put on a more
rigorous footing using the gauged linear sigma model
approach \ref\witphase{E.\ Witten, ``Phases of $N=2$ theories in
two dimensions'', {\it Nucl.\ Phys.}{\bf B403} (1993) 159.}.

Putting all this together we can conclude that the topological
theory (in the B-model) describing the conifold coupled to
topological gravity is equivalent to the $c=1$ string theory at the
self-dual radius. Since we have used ingredients from different
sources in reaching this conclusion, it would be desirable to
have a more direct check of these ideas. In the following, we will
sketch two independent arguments which further support the
above reasoning.

The first is the observation by Witten \ref\witgr{E.\ Witten,
``Ground ring of two dimensional string theory,''  {\it
Nucl.\ Phys.} {\bf B373} (1992) 187.} (see also \ref\witz{E.\ Witten and
B.\ Zwiebach,``Algebraic structures and differential geometry
in 2-d string theory'', {\it Nucl.\ Phys.}{\bf B377} (1992) 55.})
that various facts about the $c=1$ theory at the self-dual
radius with $\mu =0$ can be best described by studying the
geometry of the so called ground ring manifold $Q$ given by
$$x_1x_2-x_3x_4=0$$
The quadric cone $Q$ is precisely the conifold, provided we
interpret the variables $x_i$ as {\it complex}
coordinates\foot{This correspondence immediately suggests a
generalization of the topological model to $c=1$ string at
other special radii of compactification\ref\grad{D. Ghoshal,
work in progress.}.}. This gives us an understanding
of the main result of \witgr\witz\ which is the identification
of physical states of $c=1$ string theory with various cohomology
elements of $Q$. Indeed in a topological theory the BRST states
correspond to the cohomology of the manifold on which the sigma model
lives. In this identification, the ghost numbers of the left- and
right-moving components of the physical states are, up to shifts,
the (anti-)holomorphic degrees of the differential forms on $Q$.

Moreover the symmetries of this string theory are the `volume
preserving diffeomorphisms' of the quardric cone $Q$. This
was really established for the case of vanishing cosmological
constant. However, it was argued that the conclusions remain
unchanged for non-zero $\mu$, which has the effect of
smoothening the singularity (see \ref\glmetc{D.\ Ghoshal,
P.\ Lakdawala and S.\ Mukhi, ``Perturbations of the ground
varieties of $c=1$ string theory'', {\it Mod.\ Phys.\ Lett.}{\bf
A8} (1993) 3187; J.\ Barbon, ``Perturbing the ground ring of
2D string theory'', {\it Int.\ J.\ Mod.\ Phys.}{\bf A7} (1992) 7579;
S.\ Kachru, ``Quantum rings and recursion relations in 2D quantum
gravity'', {\it Mod.\ Phys.\ Lett.}{\bf A7} (1992) 1419.}\
for support of this argument). Now, in view of our interpretation,
the symmetry should be elevated to the `holomorphic three-form
preserving' analytic diffemorphisms of $Q$. In fact
this was one of the hints provided in \bcov\ for a connection
between topological string theory on Calabi-Yau manifolds and
the $c=1$ string.

Compared to the compact Calabi-Yau manifolds, there is however
an important difference in this case. The threefold here is
non-compact and since we have not specified the boundary conditions
the dimension of the cohomology can be infinite. This is the
reason for having infinitely many tachyons and other discrete
states in the $c=1$ string theory.

A second check follows from the observation made in \bcov\
concerning the behaviour of the singularity of partition functions
of Calabi-Yau manifolds near the conifold points. The nature of
this singularity is identical to that of the $c=1$ string at
all loops in the genus expansion of the free energy. It was
suggested that in this sense, topological strings based on
Calabi-Yau threefolds are in the same universality class as
the $c=1$ string.

However if our identification is correct not only should the
singularity structure be the same but also the coefficients
multiplying the singularity in the genus expansion of the partition
function must match exactly. This is expected since the free energy
near the conifold point is dominated by the development of the
singularity.

The $c=1$ string theory at the self-dual radius have been solved
in many different ways. For this theory, we know the exact
expression of the free energy as a function of the cosmological
constant $\mu$. This is given as a sum over world-sheet of
different genera \ref\matrix{I.\ Klebanov and D.\ Gross,``
One-dimensional string theory on a circle'', {\it Nucl. Phys.}
{\bf B344} (1990) 475.}\dv
\muv \gim:
\eqn\freen{F={1\over 2}\mu^2 \log\mu -{1\over 12} \log\mu +
{1\over 240}\mu^{-2}+\sum_{g>2} a_g \mu^{2-2g} }
where the coefficient $a_g=B_{2g}/2g(2g-2)$ is the Euler class of
the moduli space of Riemann surfaces of genus $g$.

Let us start with genus zero.
This is indeed the well-known universal behavior for the conifold.
When $\mu$ parametrizes the period of the vanishing three cycle,
there exists, in the symplectic basis, a conjugate cycle whose
period is proportional to $\mu\log\mu$. It follows from special
geometry that the free energy is given by its
integral\conifold\ref\spgeom{P.\ Candelas and X.\ de la Ossa,
``Moduli space of Calabi-Yau manifolds'', {\it Nucl.\ Phys.}
{\bf B355} (1991) 455; A.\ Strominger, ``Special geometry'',
{\it Comm.\ Math.\ Phys.}{\bf 133} (1990) 163.}. This of course
agrees with the tree level contribution to the free energy of the
$c=1$ string. It also helps us identify the normalization of
$\mu$ against the complex modulus of the degenerating threefold.

Let us now go to the one loop result. In that case, as has
recently been discussed in \vfate,
there is a lot of evidence that the coefficient of $-{1\over 12}$
accompanying $\log\mu$ is the universal behaviour of the one loop
topological partition function of the conifold. This would in fact
be necessary for the consistency of the physical resolution the
conifold singularity in string theory as proposed in \strom\gms.

This brings us to the higher loop coefficients, which would certainly
be a rather non-trivial check of our proposed link between the $c=1$
string at self-dual radius and topological theory of the conifold.
Here we notice that even though all
the higher genus free energy of the topological sigma models based
on Calabi-Yau spaces have not yet been computed, the genus-2 behaviour
of the celebrated quintic threefold has
been studied in detail in \bcov.

We should start by fixing the normalization. In the notation of
\bcov, the complex modulus $\psi$ parametrizes the behaviour of
the quintic near the conifold point $\psi=1$. The relation between
$\mu$ and $\psi$ is determined by the genus zero term as
$$\mu \sim \left({5\over 2\pi i}\right)^3 (1-\psi)$$
In \bcov, the partition function depends on $\mu$ and $\bar\mu$,
(that is why there is an `anomaly' in such theories). Setting
$\bar\mu=0$, and expanding the free energy as a function of $\mu$,
one can check that the anomaly contribution captured by the Feynman
graphs of \bcov\ does not affect the leading singularity, which
entirely comes from the coefficient $C$
(see p.\ 398--399 of \bcov)\foot{There is an implicit
${1/(2\pi i)^3}$ in the definition of $C$ as given in
in \bcov.  This arises because in defining the Yukawa couplings
on sphere
the theta angle was chosen
to be periodic with period one rather than $2\pi$.}.
Using the normalization fixed as above, the coefficient of the
$\mu^{-2}$ term is ${1\over 240}$, exactly as in the $c=1$
theory!

It would be desirable to make an explicit check of the other
higher loop coefficients. However, on the basis of our general
reasoning, we expect them to match
exactly;  in fact turning this around, the above
result provides one more data
to fix the holomorphic piece\bcov\ of the topological partition
functions at higher genera!

Thus we come to the remarkable conclusion that the genus expansion
of the free energy of the $c=1$ string theory at the self-dual
radius gives the universal singular behaviour of the free energy of
an arbitrary topological sigma model on a Calabi-Yau threefold
near a conifold point of its moduli space.

This link between the conifold and the $c=1$ string theory is sure
to have many implications. The latter is particularly well
understood from many different points of view
(for a review see \ref\ginmor{P. Ginsparg and G. Moore,
``Lectures on 2d gravity and 2d string theory'', TASI summer school,
1992; hep-th/9304011.}). Once the full
implication of the correspondence is clear, it might help understand
the structure of the conifold transition which would be important
in further developing the recent results of \strom . If so,
just as with the partition function discussed
in \vfate , for each fixed genus the tachyon scattering
amplitudes in this theory will have a bearing as exact
computations for the string theory near the conifold and should be
further reviewed in this light.  Note also that on the other hand,
as far as the $c=1$ theory is concerned, the string field
theory is thus the Kodaira-Spencer theory of \bcov\ expanded
in the conifold background, and should provide an alternative
method for computing the scattering amplitudes in this
theory\foot{Geometrically the tachyons are $(2,1)$ forms on the
threefold.}.

In \strom\gms, black holes that are solitonic solutions of the
string theory play a crucial role. The singularity
in the effective field theory near a conifold is caused by
integrating them out. Since the $c=1$ theory captures precisely
this singularity, there is probably a more direct connection of
the modes connected with these light black holes and the excitations
of the former. In this context, it is intriguing to note that
there are black hole solutions in the $c=1$ string
theory\ref\bhole{E.\ Witten, ``On string theory and black holes'',
{\it Phys. Rev.} {\bf D44} (1991)314; A.\ Sengupta,
G.\ Mandal and S.\ Wadia, ``Classical solutions of two-dimensional
string theory'', {\it Mod. Phys. Lett.} {\bf A6} (1991) 1685.}\muv.

The lesson we are learning is that we should identify the singularities
of Calabi-Yau manifolds with scaling limits of the $c=1$ string.
However there are many types of singularities of Calabi-Yau manifolds,
and for each one of them we may get a different universal theory.  Here
we have only identified the one corresponding to the simple conifold
singularity.  One might also be tempted to identify the $c=1$ string
at other special radii with degenerations of Calabi-Yau manifold.
For example at $n$ times the self-dual radius at zero cosmological
constant the expected ring is $(xy)^n=zt$ \ref\gjm{D.\ Ghoshal,
D.\ Jatkar and S.\ Mukhi, ``Kleinian singularities and the ground
ring of $c=1$ string theory'', {\it Nucl.\ Phys.}{\bf B395} (1993)
144.}. This does describe a three
dimensional non-compact singular manifold and one may wonder
whether it can arise as a part of a compact Calabi-Yau manifold.
It seems unlikely that this particular type of singularity will
appear as a part of a compact Calabi-Yau manifold for the
following reason.
The one loop partition function of this theory
$F_1=-{1\over 24}\left(n+{1\over n}\right)\log\mu$, is not,
(except for $n=1$), an integer multiple of $-{1\over 12}\log\mu$
as is required for the resolution of the singularity\vfate.

\bigskip
{\bf Acknowledgement:} We would like to thank the hospitality of the
International Centre for Theoretical Physics, Trieste, where
this work was started.  It is a pleasure to thank Michael Bershadsky,
Kirti Joshi, Shamit Kachru,  Sunil Mukhi, Hirosi Ooguri, Edward
Witten and S.-T. Yau for valuable
discussions. The research of C.V.\ was supported
in part by NSF grant PHY-92-18167.

\bigskip

\listrefs

\end